\documentclass[aps,amssymb]{revtex4}
\usepackage{amsmath}

\begin{document}

\title{Spacetimes with a vanishing second Ricci invariant}
\author{Kayll Lake}
\email{lakek@queensu.ca}

\affiliation{Department of Physics, Queen's University, Kingston, Ontario, Canada, K7L
3N6}

\begin{abstract}
Spacetimes with a vanishing second Ricci invariant, but which are not necessarily Ricci - flat, though common in general relativity, are seldom studied in a coordinate and symmetry independent way by actually using their Ricci invariants. Yet, and as an example, it can be shown by use of Ricci invariants that no such spacetimes can actually represent a perfect fluid by way of Einstein's equations. The widely studied Kiselev black hole is a particularly simple example of such a spacetime. Yet it is frequently, and erroneously, referred to as a perfect fluid. This paper gathers together information relevant to the study of spacetimes with vanishing a second Ricci invariant. It is shown that such spacetimes need not be stationary, a point relevant to their possible physical significance.
\end{abstract}
\maketitle
\section{Introduction}
Motivation for this work comes from the recent work by Visser \cite{visser1} and colleagues \cite{visser2} regarding the widely quoted Kiselev black hole \cite{kiselev}. As they make very clear, this solution is \textit{not} a perfect fluid and therefore is not related to the usual use of the term ``quintessence".
The Kiselev black hole is a particularly simple example of a spacetime with vanishing second Ricci invariant which is the subject of this brief study.
\bigskip
\section{Formalism}
Whereas the Ricci scalar ($R$) is universally accepted as the first Ricci invariant, nowadays subsequent Ricci invariants are constructed out of contractions of the trace-free Ricci tensor \cite{cm}
\begin{equation}\label{tracefree}
S^{\alpha}_{\beta} = R^{\alpha}_{\beta} -\frac{R}{4}\delta^{\alpha}_{\beta}\,,
\end{equation}
not contractions of the Ricci tensor $R^{\alpha}_{\beta}$ itself. The remaining independent Ricci invariants are given by \cite{note}
\begin{equation}\label{r1}
    r_1 =S^{\alpha}_{\beta}S^{\beta}_{\alpha}\,,
\end{equation}
\begin{equation}\label{r2}
    r_2 =S^{\alpha}_{\beta}S^{\gamma}_{\alpha}S_{\gamma}^{\beta}\,,
\end{equation}
and
\begin{equation}\label{r3}
    r_3 =S^{\alpha}_{\beta}S^{\gamma}_{\alpha}S^{\delta}_{\gamma}S^{\beta}_{\delta}\,.
\end{equation}
Ricci invariants of higher degree are not independent \cite{cmb},
\begin{equation}\label{n4}
  r_{n,n \geq 4} = r_{n}(r_{1}, r_{2}, r_{3})
\end{equation}
where $r_{n}(r_{1}, r_{2}, r_{3})$ means polynomial dependence.
The spacetimes under consideration here are characterized by the condition
\begin{equation}\label{r2zero}
    \boxed{r_2 =0}\,.
\end{equation}
It is interesting to note that (\ref{r2zero}) gives
\begin{equation}\label{reven}
    r_{n \; even} =0\,.
\end{equation}
\bigskip
\subsection{An Elementary Example}
The spacetime
\begin{equation}\label{fmetric}
ds^2 = -f(r)dt^2+\frac{dr^2}{f(r)}+r^2d\Omega^2_{2}\,,
\end{equation}
where $d\Omega^2_{2}$ is the metric of a unit two-sphere, satisfies (\ref{r2zero}) for all $f(r)$. Other properties of (\ref{fmetric}) have been studied by Jacobson \cite{jacobson}. Because of the spherical symmetry of (\ref{fmetric}), $r_3$ is not independent of $r_1$ \cite{cmb}. For (\ref{fmetric}) let us note that $r1=0$ for
\begin{equation}\label{r1zero}
f(r)=1+\frac{c_1}{r}+c_2r^2
\end{equation}
and $R=0$ for
\begin{equation}\label{Rzero}
f(r)=1+\frac{c_1}{r}+\frac{c_2}{r^2}\,,
\end{equation}
where $c_1$ and $c_2$ are constants. These forms of $f(r)$ are very familiar as regards solutions within the context of Einstein's equations \cite{solutions} to which we now turn.
\section{Interpretation}
Einstein's equations are
\begin{equation}\label{einstein}
   R^{\alpha}_{\beta}-\frac{1}{2}\delta^{\alpha}_{\beta}R+\Lambda \delta^{\alpha}_{\beta} = 8 \pi T^{\alpha}_{\beta}
\end{equation}
where $\Lambda$ the cosmological constant and $T^{\alpha}_{\beta}$ the energy - momentum tensor \cite{em}. We use these equations to interpret the Ricci invariants in terms of the energy - momentum tensor. The Ricci scalar follows as
\begin{equation}\label{ricci}
    R=4\Lambda-8 \pi T
\end{equation}
where $T$ is the trace of $T^{\alpha}_{\beta}$. Einstein's equations now take the convenient form
\begin{equation}\label{traceeinstein}
    S^{\alpha}_{\beta} = 8 \pi (T^{\alpha}_{\beta}-\frac{T}{4}\delta^{\alpha}_{\beta})\,.
\end{equation}
From (\ref{traceeinstein}) we have the two identities \cite{8pi}
\begin{equation}\label{tr1}
r_1 = T^\alpha_\beta T^\beta_\alpha-\frac{T^2}{4}
\end{equation}
and
\begin{equation}\label{tr2}
r_2 = T^\alpha_\beta T^\beta_\gamma T^\gamma_\alpha-\frac{3}{4}T T^\alpha_\beta T^\beta_\alpha+\frac{T^3}{8}
\end{equation}
so that for the spacetimes under consideration here we must have
\begin{equation}\label{ttr2}
  T^\alpha_\beta T^\beta_\gamma T^\gamma_\alpha+\frac{T^3}{8} = \frac{3}{4}T T^\alpha_\beta T^\beta_\alpha\,.
\end{equation}
\subsection{Perfect Fluids}
Consider a perfect fluid so that
\begin{equation}\label{perfect}
T_{\alpha}^{ \beta} = (\rho + p) u_{\alpha} u^{\beta} + p \delta_{\alpha}^{ \beta}
\end{equation}
where $\rho$ is the energy density, $p$ is the (isotropic) pressure, and
$u^{\alpha}$ is the normalized timelike flow vector.  The Ricci invariants in this case \textit{all}
reduce to the form
\begin{equation}
r_{n}=(\rho + p)^{n+1}\,,
\end{equation}
where again $=$ means equivalence up to a physically irrelevant constant. From (\ref{r2zero}) then the spacetimes under consideration here, assuming a perfect fluid decomposition of the energy - momentum tensor, are invariantly characterized by
\begin{equation}\label{rhopzero}
r_{n}=\rho+p=0\,.
\end{equation}
For example, for spacetimes of the form (\ref{fmetric}), (\ref{r1zero}) must hold and so we have only the familiar Kottler metric.
\bigskip
\subsection{Another Elementary Example}
The spacetime
\begin{equation}\label{fwmetric}
ds^2 = 2 \epsilon dr dw+r^2d\Omega^2_{2}-f(r,w) dw^2\,,
\end{equation}
where $\epsilon^2 = 1$, satisfies (\ref{r2zero}) for all $f(r,w)$. Because of the spherical symmetry $r_3$ is again not independent of $r_1$. For (\ref{fwmetric}) let us note that $r_1=0$ for
\begin{equation}\label{r1wzero}
f(r,w)=1+\frac{c_1(w)}{r}+c_2(w)r^2
\end{equation}
and $R=0$ for
\begin{equation}\label{Rwzero}
f(r,w)=1+\frac{c_1(w)}{r}+\frac{c_2(w)}{r^2}\,.
\end{equation}
These forms of $f(r,w)$ are familiar generalizations of the Vaidya metric \cite{solutions}. What is of interest here is the fact that the metric (\ref{fwmetric}) is not stationary unless $\partial f(r,w)/ \partial w = 0$. This expands the possible physical relevance of spacetimes that satisfy (\ref{r2zero}).

\bigskip
\textit{Acknowledgments.} This work was supported by a grant from the Natural Sciences and Engineering Research Council of Canada. This work was made possible by the use of GRTensor \cite{grt}.


\begin{thebibliography}{}
\bibitem{note}Physically irrelevant numerical coefficients arise due to the translation from spinor notation to tensor notation. These coefficients are simply ignored here.
\bibitem{em}No decompostion of the energy - momentum tensor is considered here.
\end{thebibliography}
\end{document}